\documentclass[reqno]{amsart}
\usepackage{amssymb,stmaryrd}
\usepackage{amsfonts}
\usepackage{amstext}
\usepackage{algorithmic}
\usepackage{algorithm}
\usepackage{color}
\usepackage{graphicx}
\usepackage{epstopdf}
\usepackage[all]{xy}
\usepackage{MnSymbol, slashed}

\numberwithin{equation}{section}
\newtheorem{theorem}{Theorem}[section]
\newtheorem{lemma}[theorem]{Lemma}
\newtheorem{proposition}[theorem]{Proposition}

\theoremstyle{definition}

\theoremstyle{remark}

\newcommand{\C}{{\mathbb{C}}}

\newcommand{\<}{{\langle}}

\renewcommand{\>}{{\rangle}}

\newcommand{\CH}{{\mathcal{H}}}

\newcommand{\CS}{{\mathcal{S}}}

\newcommand{\CJ}{{\mathcal{J}}}
\newcommand{\cg}{{\mathfrak{g}}}
\newcommand{\wedgeq}{{\wedge\kern-5pt\cdot}}

\newcommand{\tens}{\otimes}

\newcommand{\id}{{\rm id}}

\newcommand{\extd}{{\rm d}}
\newcommand{\del}{{\partial}}
\newcommand{\eps}{\epsilon}

\newcommand{\la}{{\triangleright}}

\newcommand{\dirac}{{ \slashed{D} }}

  \allowdisplaybreaks

\begin{document}

\title{Fermions in the fuzzy sphere Kaluza-Klein model }


\author{Chengcheng Liu and Shahn Majid}

\address{Queen Mary University of London\\
School of Mathematical Sciences, Mile End Rd, London E1 4NS, UK}

\email{s.majid@qmul.ac.uk}
\thanks{The first author was supported by a China Scholarship}

\begin{abstract} We consider spinors on the total space of a Kaluza-Klein model with fuzzy sphere fibre and geometrically realised Dirac operator on the product. We show that a single massless spinor on the product appears on spacetime as multiplets of spinors with a particular signature of differing masses and $SU(2)$ Yang-Mills charges. For a finite-dimensional fibre, these become finite multiplets with different masses rather than infinite towers. For example, for the reduced fuzzy sphere isomorphic to $M_2(\C)$, a massless spinor appears as two $SU(2)$ doublets and an $SU(2)$ quadruplet in mass ratios $1:5/3:7/3$.  Although such signatures do not match known fermions in the Standard Model, the paper provides  a new mechanism which could be further explored for other noncommutative fibre algebras. \end{abstract}
\maketitle 

\section{Introduction}

In recent works \cite{LiuMa2, LiuMa3}, we have revisited the Kaluza-Klein idea but with the `compact fibre' at each point of spacetime replaced by a fibre with a  noncommutative coordinate algebra, i.e. using noncommutative geometry. The key difference is that while in conventional Kaluza-Klein theory one has to make a `cylinder ansatz' of a very special form of metric on the product in order to encode gravity and Yang-Mills on spacetime, that ansatz turns to be  exactly what is forced by the noncommutative geometry under the minimal assumptions of (a) a tensor product differential structure (so that the classical spacetime differential forms and fibre quantum differential forms commute with each other) and (b) the fibre quantum geometry having a central basis of 1-forms and the fibre coordinate algebra having trivial centre. The latter are quite typical for highly noncommutative quantum geometries. As a result, one can expect somewhat generally, i.e. not only for the specific fuzzy sphere fibre in \cite{LiuMa2}, that gravity on the product space literally decomposes as gravity on spacetime, Yang-Mills on spacetime and a positive-matrix valued Liouville/sigma model field on spacetime  for the metric on the fibre as it varies over spacetime. We also looked at how a scalar field on the product appears as a multiplet of fields on spacetime of varying Yang-Mills charges and masses. 

In this 3rd work in the series, we examine the physically more relevant case of how a spinor or fermion field on the product appears as a multiplet of spinors on spacetime, finding for the fuzzy sphere a very particular signature of Yang-Mills charges and masses. Our goal is to provide proof of concept that the mechanism works and could in principle be applied to other quantum fibre geometries with potential implications for particle physics. This therefore completes a potential new mechanism for why particular matter and gauge fields observed on spacetime could be forced out of gravity on the product and the idea that geometry at the Planck scale could be noncommutative as in \cite{Sny,Hoo,Ma:pla,DFR}. That the fibre should in this case be `quantum' is quantitatively supported by the calculation (with similar results even in conventional Kaluza-Klein theory with compact fibre) that the size of the fibre geometry, to match the observed coupling constants for Maxwell and electro-weak Yang-Mills, needs to be around 11 Planck lengths / 23  Planck lengths respectively, see \cite{LiuMa2,LiuMa3}. We will also consider the possibility of a much weaker gauge field on spacetime  for a different as yet unobserved or approximate $SU(2)$ gauge symmetry. In this case, the fibre geometry scale would be much larger but we can still posit it as noncommutative even if this noncommutativity does not arise as a quantum gravity correction. A further  feature of the noncommutative fibre approach is that we can also take the coordinate algebra on the fibre to be finite-dimensional, giving finite multiplets rather and infinite towers of fields as in regular Kaluza-Klein theory. 

Turning to the formalism, the idea of using a noncommutative fibre in this context goes back to  pioneering work of Connes and collaborators, notably \cite{ChaCon} under the heading of `almost commutative' geometry. We adopt the same setup where the product geometry is described by an algebra $A=C^\infty(M)\tens A_f$, where $M$ is spacetime and $A_f$ is a noncommutative fibre coordinate algebra. We will take $A_f=\C_\lambda[S^2]$ the fuzzy sphere  or, for finite multiplets,  $A_f=c_\lambda[S^2]$ the reduced fuzzy sphere. In the Connes approach, one works directly with the `Dirac operator' $\dirac$ defined abstractly by certain axioms as a `spectral triple' $(A,\dirac,\CH)$, where $\CH$ is a Hilbert space, in the product case built from spinor fields on $M$ tensored with a representation space for $A_f$. Moreover, $\dirac$ is a `tensor product' of $\dirac_M$ on spacetime and $\dirac_f$ for the fibre (there is an extra $\gamma_5$ in front of the latter). While the Connes approach remains very much of interest in relation to particle physics, e.g.\cite{DabSit}, it does not seek to build the total space $\dirac$ operator and spinor fields directly in a fully geometric manner, for example $\dirac_f$ is not associated to (fibre) spinor bundle and spin connection. In this context, a more geometric framework to construct Dirac operators on noncommutative algebras $A$ was proposed in \cite{BegMa:spe} using a more explicit `quantum Riemannian geometry' (QRG) approach which starts with a differential exterior algebra $(\Omega,\extd)$ on $A$, constructs a quantum metric $\cg\in \Omega^1\tens_A\Omega^1$ and a `quantum Levi-Civita connection' (QLC) $\nabla:\Omega^1\to \Omega^1\tens_A\Omega^1$. Then, in this context, we seek to construct a spectral triple in the form $\dirac=\la\circ\nabla$  using the further geometric data of a spinor bundle $\CS$, spinor connection $\nabla_\CS:\CS\to\Omega^1\tens \CS$ on it, a `Clifford action' $\la: \Omega^1\tens_A\CS\to \CS$ and certain other data, notably a charge conjugation operator. The geometry applies at a `local tensorial' level before we consider an inner product $\<\ ,\ \>$ on $\CS$. Including that but not yet completion to a Hilbert space is what we call a `pre-spectral triple' or pre-Hilbert space level, and ideally we would then want to complete this to an actual  spectral triple. This is closer to what is familiar in classical geometry and includes  additional rules (not needed for Connes axioms) that link $\nabla_\CS$ with $\nabla$ and the Clifford action with the quantum metric.  Also, by not directly going to the operator algebras and that style of noncommutative geometry\cite{Con}, we can look more freely to Lorentzian signatures where the precise axioms of a Connes spectral triple are not yet clear, albeit there are some ideas such as in \cite{Dev}. 

Section~\ref{secpre} provides a recap of the formalism of QRG, as applied in physics in works such as \cite{Ma:sq,Ma:haw, LirMa2, LiuMa1}, and in particular the geometric realisation formalism for spectral triples. The latter was recently applied to noncommutative algebras in \cite{LirMa2,LirMa3,Ma:dir}, while Section~\ref{sec:class} explains in detail how the formalism is realised for $C^\infty(M)$ in the standard treatment of spinors in the Physics literature.  More details for the general formalism can be found in \cite{BegMa}.  Section~\ref{sec:KK} then proceeds to the new results of the paper, where we apply the geometric realisation formalism for Dirac operators/spectral triples to the product $A=C^\infty(M)\tens A_f$, focussing on $A_f$ again a fuzzy sphere. Section~\ref{sec:multi} then interprets how spinors on the product look from the point of view of multiplets of spinors on spacetime. Some of this, namely the change of variables to turn a chiral mass into an ordinary one, is adapted from the treatment of spinors in classical Kaluza-Klein theory \cite{freedman}. The paper ends in Section~\ref{sec:con} with some concluding remarks about the physical interpretation and values of the parameters in the model.

\section{Preliminaries on quantum Riemannian geometry and spectral triples}\label{secpre}

Here, we explain the basic elements of the noncommutative geometry in the form that we need, including how the conventional treatment of Dirac operators on a manifold in the literature  appears from our point of view. Then, in later sections, we will apply the formalism to  the tensor product algebra $A=C^\infty(M)\tens A_f$ where $A_f$ has  its own geometrically realised spectral triple or `Dirac operator'.

\subsection{Recap Connes spectral triples}\label{RealS}  

We first recall the definition of a real spectral triple $(A,\CS,\<\ , \ \>, \dirac,\CJ,\gamma)$ as in \cite{ChaCon} and elsewhere. This is such that:

(1) $\CS$ is a Hilbert space with inner product $\<\ ,\ \>$ on which a $*$-algebra $A$ acts faithfully with the action of $a$* adjoint to the action of $a$, and on which operators $\dirac,\CJ$ and in the `even' case $\gamma$ act. Moreover, $\dirac$ is antihermitian(which in Connes' convention is replaced by hermitian $D=-\imath\dirac$), $\CJ$ is an antilinear isometry and $\gamma$ is hermitian.

(2)\ $[a,\CJ b\CJ^{-1}]=[[\dirac,a],\CJ b\CJ^{-1}]=[\gamma,a]=0$ for all $a,b\in A$.

(3)\ $\CJ^2=\epsilon\id,\ \CJ\dirac=\epsilon'\dirac\CJ,\ \CJ\gamma=\epsilon''\gamma\CJ,\ \gamma^2=\id,\ \dirac\gamma=-\gamma\dirac$.

The  signs $\eps,\eps',\eps''$ are normally given by the period 8 table (with $\eps'$ the negative of $\eps'$ in Connes papers due to the different convention),

\[ \begin{array}{c|c|c|c|c|c|c|c|c}n& \ \ \  0 \ \ \  & \ \ \  1 \ \ \  & \ \ \  2 \ \ \  & \ \ \  3 \ \ \  & \ \ \  4 \ \ \  & \ \ \  5 \ \ \  & \ \ \  6 \ \ \  & \ \ \  7 \ \ \  \\ \hline
\epsilon\phantom{\Big|}& 1 & 1 & -1 & -1& -1& -1 & 1& 1\\ \hline 
\epsilon'\phantom{\Big|}& -1 & 1 & -1 & -1& -1& 1 & -1& -1\\ \hline
\epsilon''\phantom{\Big|}& 1 &  & -1 & & 1&  & -1& \\ \hline
\end{array}\]
but we are not limited to this table.

\subsection{Recap of geometric realisation in QRG} \label{GS}
 We work with $A$ a unital algebra, typically a $*$-algebra over $\C$, in the role of `coordinate algebra'. Differentials are formally introduced as an $A$-bimodule $\Omega^1$ of 1-forms equipped with a map $\extd:A\to \Omega^1$ obeying the Leibniz rule 
\[ \extd(ab)=(\extd a)b+a\extd b\]
for all $a,b\in A$. This is required to extend to an exterior algebra $(\Omega,\extd)$ generated by $A,\extd A$, with $\extd^2=0$ and $\extd$ obeying the graded-Leibniz rule.   The exterior algebra is called {\em inner} if there exists $\theta\in\Omega^1$ such that $\extd=[\theta,\ \}$ is a graded derivation. We also say that $\Omega^1$ by itself is inner if $\extd=[\theta,\ ]$ acting on $A$.

A quantum metric is $\cg\in \Omega^1\tens_A\Omega^1$ together with a bimodule map inverse $(\ ,\ ):\Omega^1\tens_A\Omega^1\to A$ in the sense 
\[( (\omega,\ )\tens\id)\cg=\omega=(\id\tens (\ ,\omega))\cg\]
for all $\omega\in \Omega^1$, and some form of quantum symmetry condition such as $\wedge(\cg)=0$ (we refer to $\cg$ as a generalised quantum metric if no form of symmetry is imposed).  A (left) bimodule connection\cite{DVM,Mou} on $\Omega^1$ is $\nabla:\Omega^1\to \Omega^1\tens_A\Omega^1$ obeying 
\[\nabla(a \omega)=\extd a\tens\omega+a\nabla\omega,\quad \nabla(\omega a)=\sigma(\omega\tens \extd a)+(\nabla \omega)a\]
for all $a\in A,\omega\in \Omega^1$ where $\sigma:\Omega^1\tens_A \Omega^1\to \Omega^1\tens_A \Omega^1$ is a bimodule map. A connection is {\em metric compatible} if the tensor product connection
\[\nabla_{\Omega^1\tens\Omega^1}:=\nabla\tens\id+(\sigma\tens\id)(\id\tens\nabla)\]
vanishes on $\cg$. This can also be said in terms of $(\ ,\ )$. 

The formulation of a bimodule connection above applies equally well to any $A$-bimodule $\CS$ in the role of sections of a `vector bundle', now with $\nabla_\CS: \CS\to\Omega^1\tens_A \CS$, $\sigma_\CS:\CS\tens_A\Omega^1\to\Omega^1\tens_A\CS$ and $\nabla_{\Omega^1\tens \CS}:\Omega^1\tens \CS\to\Omega^1\tens\Omega^1\tens \CS$ with
\[\nabla_\CS(a \psi)=\extd a\tens\psi+a\nabla_\CS\psi,\quad \nabla_\CS(\psi a)=\sigma_\CS(\psi\tens \extd a)+(\nabla_\CS \psi)a,\]
\[\nabla_{\Omega^1\tens \CS}=\nabla\tens\id+(\sigma\tens\id)(\id\tens\nabla_\CS)\]
for all $a\in A,\psi\in \CS$.

Next, we introduce $\la:\Omega^1\tens_A\CS\to\CS$ as a {\em Clifford bimodule map}, which generalises the role of the gamma-matrices when constructing Dirac operators, and define $\dirac=\la\circ\nabla_\CS: \CS\to \CS$ as a minimal `geometric Dirac operator'. Then a spectral triple is `geometrically realised'\cite{BegMa:spe} at the local tensorial level if $\CJ,\la,\dirac$ obey following conditions. 
 (i) $\CJ:\CS\to\CS$ is an antilinear skew-bimodule map in the sense
 \begin{align}\label{1}
 \CJ(a\psi)=\CJ(\psi)a^*,\quad\CJ(\psi a)=a^*\CJ(\psi),\quad \CJ^2=\epsilon\id
 \end{align}
for all $a\in A,\psi\in \CS$ and some $\epsilon=\pm 1$. (ii) $\nabla_\CS,\CJ$ obey
\begin{align}\label{2}
\nabla_\CS\CJ=\sigma_\CS\circ {\rm flip}(*\tens\CJ)\circ\nabla_\CS,\quad \CJ(\omega\la\psi)=\epsilon'\la\circ\sigma_\CS(\CJ\psi\tens\omega^*)
\end{align}
for all $\omega\in\Omega^1,\psi\in\CS$.  (iii) for an even spectral triple, there is a bimodule map $\gamma:\CS\to\CS$ with $\gamma^2=\id$ required to further obey
\begin{align}\label{3}
(\id\tens\gamma)\circ\nabla_\CS=\nabla_\CS\circ\gamma,\quad\gamma\circ\la=-\la\circ(\id\tens\gamma),\quad \CJ\circ\gamma=\epsilon''\gamma\circ\CJ.
\end{align}
In this way, we naturally realise the local tensorial part of Connes axioms of a spectral triple. Next we need $\<\ ,\ \>$ a sesquilinear inner product on $\CS$ to form a Hilbert space such that $\gamma$ is hermitian, $\dirac$ is antihermitian, $\CJ$ an antilinear isometry in the sense
\[\<\CJ\phi,\CJ\psi\>=\<\psi,\phi\>\]
for all $\psi,\phi\in\CS$. For an actual Connes spectral triple one then requires the Dirac operator to have bounded commutators and a compact resolvent. 

\subsection{Full geometric realisation with respect to a QRG} Although not needed for a special triple, ideally we would also like a {\em full geometric realisation} in the sense\cite{BegMa:spe} 
\begin{equation}\label{covcliff}  (\id\tens\la)\circ\nabla_{\Omega^1\tens \CS}=\nabla_\CS\circ\la,\end{equation}
which says that $\la$ is covariantly constant (it intertwines $\nabla_{\Omega^1\tens \CS}$ and $\nabla_\CS$). \cite{BegMa:spe} also tentatively proposed a `Clifford relation' that  $\la$ extends to $\Omega^2\tens_A\CS\to \CS$ in such a way that 
\begin{equation}\label{cliffreln} (\omega\wedge\eta)\la \phi=\omega\la(\eta\la \phi)-(\omega,\eta)\phi\end{equation}
for all $\omega,\eta\in \Omega^1$ and $\phi\in \CS$, or some variant of this condition. The motivation here is that this reduces to the Clifford algebra relations in the classical case. Both of these geometric conditions are part of the quantum Riemannian geometry and {\em not} part of the requirements for $\dirac$ to be part of a Connes spectral triple.

\subsection{Classical Dirac operator on $M$ in spectral triple terms}\label{sec:class}

We let $e^A$ be (local) basis of the spinor bundle and $\tilde g^{\mu\nu}$ a metric on $n$-dimensional (pseudo)-Riemannian manifold $M$. We introduce $n$-beins $\{e^{a}\}$ such that
\[\tilde g_{\mu\nu}\extd x^\mu\tens\extd x^\nu=\eta_{ab}e^{a}\tens e^{b}\]
and define $e_{a}^{\ \mu},e_{\mu}^{\ a}$ by
\[\extd x^\mu=e_{a}^{\ \mu}e^{a},\quad  e_{a}^{\ \mu}e_{\mu}^{\ b}=\delta^{\ b}_{a},\]
 from which one can easily show that
 \[\tilde g^{\mu\nu}=\eta^{ab}e_{a}^{\ \mu}e_{b}^{\ \nu},\quad\tilde g_{\mu\nu}=\eta_{ab}e_{\mu}^{\ a}e_{\nu}^{\ b},\quad e_{\nu}^{\ a}e_{a}^{\ \mu}=\delta^{\ \mu}_{\nu}.\]
 We let 
\[ \nabla_\CS e^{A}=(S_\mu)_{B}{}^{A}\extd x^\mu\tens e^{B},\quad \extd x^\mu \la e^{A}=(\tilde\gamma^\mu){}_{B}{}^Ae^{B},\quad \CJ(e^{A})=(J_M)_{B}{}^A e^{B},\quad \gamma(e^{A})=\kappa_{B}{}^A e^{B}\]
be the data for the spectral triple at the local tensorial level in terms of matrices. We further define   $\tilde\gamma^\mu=\gamma^a e_a^{\ \mu}$ in terms the $n$-bein and set
\begin{align}\label{gmma}
S_{\mu}=\frac{1}{4}(\omega_{ab})_\mu\gamma^{ab},\quad \{\gamma^{a},\gamma^{b}\}=2\eta^{ab}\id, 
\end{align}
where
\[(\omega_{ab})_\mu=\tilde g_{\rho\sigma}e^{\ \rho}_{a}\tilde \nabla_\mu e^{\ \sigma}_{b},\quad  \gamma^{ab}=\frac{1}{2}[\gamma^{a},\gamma^{b}]\]
so that
\begin{align}
\dirac (\Psi_{A} e^{A})&=\Big(\tilde\dirac_M\Psi\Big)_{A} e^{A},\quad \tilde\dirac_M=\tilde\gamma^\mu\Big(\del_\mu+\frac{1}{4}(\omega_{a b})_\mu\gamma^{ab}\Big).
\end{align}
Here, $\Psi$ denotes the column vector consisting of the elements $\Psi_{A}$.

Following some relatively standard Physics conventions in \cite{freedman}, we choose a hermitian representation of the Clifford algebra in the sense
\begin{equation}\label{g0daggger} \gamma^{a\dagger}=\gamma^{0}\gamma^{a}\gamma^{0}\end{equation}
which implies
\begin{align}\label{ts5}
\gamma^{0\dagger}=-\gamma^0,\quad (\gamma^{a}\gamma^{0})^\dagger=\gamma^{a}\gamma^{0},\quad (\gamma^{ab}\gamma^{0})^\dagger=\gamma^{ab}\gamma^{0},\quad\kappa^{\dagger}=\kappa,\quad (\kappa\gamma^{0})^\dagger=\kappa\gamma^{0}.
\end{align}
Here, in the case of even $n$, we define $\kappa_M$  as the  highest rank tensor element of the Clifford algebra in the representation,
 \begin{align}\label{JM2}
\kappa_M=(-\imath)^{{n\over 2}+1}\gamma^0\gamma^1...\gamma^{n-1}.
\end{align}
Note that the first of (\ref{ts5}) immediately implies 
\begin{align}\label{ts2}
(\tilde\gamma^\mu\gamma^{0})^\dagger=(\gamma^{a}e_{a}^{\ \mu}\gamma^{0})^\dagger=e_{a}^{\ \mu}(\gamma^{a}\gamma^{0})^\dagger=e_{a}^{\ \mu}\gamma^{a}\gamma^{0}=\tilde\gamma^\mu\gamma^{0}
\end{align}
as noted in \cite{KKhermitian}. Furthermore, it is usual to consider a charge conjugation operator satisfying
\[CC^\dagger=\id,\quad C^{\intercal}=-\epsilon_M\epsilon_M ' C,\quad \gamma^{a}{}^\intercal=-\epsilon_M ' C\gamma^a C^{-1}\]
for some  choice of signs and an appropriate representation of the Clifford algebra. This corresponds for our spectral triple data to
\[ J_M=\gamma^0 C^*\]
which then defines $\CJ_M$ as an antilinear skew-bimodule map. The properties of $C$ translate to 
  \begin{align}\label{JM}
  J_M J_M^*=\epsilon_M\id,\quad \gamma^{a *}=\epsilon_M' J_M^{-1}\gamma^a J_M.
 \end{align}
 Note that in our conventions, due to the properties of $C$, $J_M$ also obeys
\begin{align}\label{JJ1}
J_M^\dagger J_M=\id
\end{align}
as a matrix.  The signs for the spectral triple at this level and in Physics conventions for a Lorentzian metric are given by the following period 8 table 
\[ \begin{array}{c|c|c|c|c|c|c|c|c}n& \ \ \  0 \ \ \  & \ \ \  1 \ \ \  & \ \ \  2 \ \ \  & \ \ \  3 \ \ \  & \ \ \  4 \ \ \  & \ \ \  5 \ \ \  & \ \ \  6 \ \ \  & \ \ \  7 \ \ \  \\ \hline
\epsilon_M\phantom{\Big|}& \pm 1 & 1 & 1 & 1& \pm 1& -1 & -1& -1\\ \hline 
\epsilon_M '\phantom{\Big|}& \mp 1 & -1 & \pm 1 & 1& \pm 1& -1 & \pm 1& 1\\ \hline
\end{array}\]
For example, in four dimensions $\epsilon_M=\epsilon_M '=\pm 1$  (there are two choices). Moreover, for a spectral triple, the last part of (\ref{3}) translates to 
\begin{equation}\label{kappa*} \kappa_M^{*}=(-1)^{{n\over 2}+1}\epsilon_M''\kappa_M\end{equation}
for some further choice of sign $\eps_M''$. This again depends on the  Clifford algebra representation and typically holds with different choices of sign depending on the signature and type of spinor. 

Next, we define the inner product \[\left\langle \psi_{A}e^{A},\phi_{B}e^{B}\right\rangle=\int \overline\psi\phi,\]
where $\overline{\psi}=-\imath\psi^{\dagger}\gamma^0$. With respect to this inner product, $\dirac_M$ is hermitian in the sense
\begin{align*}
\langle \dirac (\psi_{A}e^{A}),\phi_{B}e^{B}\rangle&=-\imath\int\Big(\tilde\dirac_M\psi\Big)^\dagger\gamma^0\phi=\imath\int\Big(\gamma^0\tilde\dirac_M\psi\Big)^\dagger\phi\\
&=\imath\int\Big(\gamma^0\tilde\gamma^\mu\Big(\tilde\nabla_\mu+\frac{1}{4}(\omega_{a b})_\mu\gamma^{ab}\Big)\psi\Big)^\dagger\phi\\
&=\imath\int\Big(\gamma^0\Big(-\tilde\nabla_\mu \tilde\gamma^\mu+\frac{1}{4}(\omega_{a b})_\mu \tilde\gamma^\mu\gamma^{ab}\Big)\psi\Big)^\dagger\phi-\imath\int\Big(\gamma^0\tilde\gamma^\mu\psi\Big)^\dagger\tilde\nabla_\mu\phi\\
&=\imath\int\Big(\gamma^0\Big(\frac{1}{4}(\omega_{a b})_\mu \gamma^{ab}\tilde\gamma^\mu\Big)\psi\Big)^\dagger\phi-\imath\int\Big(\gamma^0\tilde\gamma^\mu\psi\Big)^\dagger\tilde\nabla_\mu\phi\\
&=-\imath\int\psi^\dagger \gamma^0\Big(\frac{1}{4}(\omega_{a b})_\mu \tilde\gamma^\mu\gamma^{ab}\Big)\phi-\imath\int\psi^\dagger\gamma^0\tilde\gamma^\mu\tilde\nabla_\mu\phi\\
&=-\imath\int \psi^\dagger\gamma^0\tilde\dirac_M\phi=\int \overline\psi\tilde\dirac_M\phi=\langle \psi_{A}e^{A},\dirac(\phi_{B}e^{B})\rangle
\end{align*}
where we used $\gamma^{0\dagger}=-\gamma^0$ for the second equality, dropped a total derivative of $\tilde\nabla_\mu$ for the fourth equality, $\tilde\nabla_\mu\tilde\gamma^\nu-\frac{1}{4}(\omega_{ab})_{\mu}[\tilde\gamma^\nu,\gamma^{ab}]=0$ for the fifth equality, and the hermitian conditions (\ref{ts5}),(\ref{ts2}) for the sixth equality.

We also calculate
\begin{align*}
&\<\CJ(\phi_{A}e^{A}),\CJ(\psi_{B}e^{B})\>=\<(J_M)^{\ A}_{C}\phi_{A}^*e^{C},(J_M)^{\ B}_{D}\psi_{B}^*e^{D}\>\\
&=\int \overline{J_M\phi^*}J_M\psi^*=-\imath\int (J_M\phi^*)^\dagger\gamma^0 J_M\psi^*=-\imath\int \phi^{\intercal} J_M^\dagger\gamma^0 J_M\psi^*\\
&=-\imath\epsilon'_M\int \phi^{\intercal} \gamma^{0*}\psi^*=-\imath\epsilon'_M\int (\phi^{\intercal} \gamma^{0*}\psi^*)^{\intercal}=\imath\epsilon'_M\int \psi^\dagger\gamma^{0}\phi=-\epsilon'_M\int \overline\psi\phi,
\end{align*}
where we used the second condition of (\ref{JM}) and (\ref{JJ1}) for the fifth equality. In addition,
\begin{align*}
\langle \gamma &(\psi_{A}e^{A}),\phi_{B}e^{B}\rangle=\langle (\kappa_M)^{\ A}_C \psi_{A} e^{C},\phi_{B}e^{B}\rangle=\int\overline{\kappa_M \psi}\phi=-\imath\int(\kappa_M\psi)^\dagger\gamma^0\phi\\
&=-\imath\int\psi^\dagger\kappa_M\gamma^0\phi=\imath\int\psi^\dagger\gamma^0\kappa_M\phi=-\int\overline{\psi}\kappa_M\phi=-\langle\psi_{A}e^{A},\gamma(\phi_{B}e^{B})\rangle.
\end{align*}
We see that $\CJ$ is formally an isometry or `anti-isometry' depending on dimension and $\gamma$ is `anti-self-adjoint', which is different from a usual spectral triple. Moreover, the inner product is not positive, so we do not have a Hilbert space. 

\section{Noncommutative Kaluza-Klein set up with spinors}\label{sec:KK}

In this section, we first look at the general set up for geometrically realised spectral triples on a product $A=C^\infty(M)\tens A_f$. We then specialise to calculations in the case where $A_f$ is a fuzzy sphere. 

\subsection{Geometrically realised spectral triples on the product}

We next explain, given the classical Dirac operator on a classical $n$-dimensional Lorentzian spacetime $M$ as in Section~\ref{sec:class} and a finite geometrically realised one on $A_f$, what is the form of a geometrically realised Dirac operator on $A=C^\infty(M)\tens A_f$ with $\Omega^1=\Omega^1(M)\tens A_f\oplus C^\infty(M)\tens \Omega^1_f$ and $\CS=\CS_M\tens \CS_f$. We suppose $\Omega^1_f$ has a central basis $s^i$ for $i=1,\cdots,n_f$ and $\CS_f$ a central basis $e^\alpha$. On $M$ we assume locally $e^A$ a (central) basis of $\CS_M$ and  $\extd x^\mu$ a (central) basis of $\Omega^1_M$. Then  we can write the central basis of $\Omega^1$ as given by  $\extd x^\mu\tens 1, 1\tens s^i$ and the central basis of $\CS$ by $e^A\tens e^\alpha$. For simplicity, we redefine $\extd x^\mu\equiv\extd x^\mu\tens 1,s^i\equiv 1\tens s^i$ and denote them by $\omega^I$, where $\omega^I$ is $\extd x^\mu,s^i$ when $I$ is $\mu,i$ respectively, and we write $e^{A\alpha}\equiv e^A\tens e^\alpha$. 
 
\begin{proposition}\label{local} If $A_f$ has trivial centre then $\nabla_{S}$, $\CJ$, $\la$ taken in the general form
\begin{align*}
&\nabla_\CS e^{A\alpha}=(S_I)_{B\beta}^{\ \ \ A\alpha}\omega^I\tens e^{B\beta},\quad \sigma_\CS(e^{A\alpha}\tens \omega^I)=\omega^I\tens e^{A\alpha},\\
\omega^I\la e^{A\alpha}&=(\gamma^I)^{\ \ \  A\alpha}_{B\beta}e^{B\beta},\quad
 \CJ(e^{A\alpha})=(J)^{\ \ \ A\alpha}_{B\beta}e^{B\beta},\quad \gamma(e^{A\alpha})=\kappa^{\ \ \ A\alpha}_{B\beta} e^{B\beta}
\end{align*}
must have coefficients in $C^\infty(M)$. Moreover, they obey the conditions for a spectral triple at the local tensorial level iff 
\begin{align}\label{tq2}
JJ^*=\epsilon\id,\quad \del_I J+S_I J=J S_I^*,\quad J\gamma^{I*}=\epsilon'\gamma^I J,
\end{align}
where $*$ means the complex conjugation of the matrix entries, and in the even case
\begin{align}\label{tq3}
\del_I \kappa+S_I\kappa=\kappa S_I,\quad \kappa\gamma^I+\gamma^I\kappa=0,\quad J\kappa=\epsilon''\kappa J,\quad\kappa^2=\id. 
\end{align}
The resulting Dirac operator $\dirac$ acting on $\Psi_{A\alpha} e^{A\alpha}\in S$ is 
\begin{align}\label{dirac3}
&\dirac (\Psi_{A\alpha} e^{A\alpha})=(\gamma^I\del_I+\gamma^I S_I)^{\ \ \ B\beta}_{A\alpha}\Psi_{B\beta}e^{A\alpha}=((\gamma^I\del_I+\gamma^I S_I)\Psi)_{A\alpha}e^{A\alpha}, 
\end{align}
where $\Psi$ denotes the column vector consisting of the elements $\Psi_{A\alpha}$.
\end{proposition}
\proof 
According to centrality of the basis and the flip map $\sigma_S$, we note
\begin{align*}
(\del_I a)&\omega^I\tens e^{A\alpha}+a\nabla_\CS e^{A\alpha}=\extd a\tens e^{A\alpha}+a\nabla_\CS e^{A\alpha}=\nabla_\CS(a e^{A\alpha})\\
=&\nabla_\CS(e^{A\alpha} a)=(\del_I a)\sigma_\CS(e^{A\alpha}\tens \omega^I)+(\nabla_\CS e^{A\alpha})a=(\del_I a)\omega^I\tens e^{A\alpha}+(\nabla_\CS e^{A\alpha})a
\end{align*}
or
\[a(S_I)^{\ \ \ A\alpha}_{B\beta}=(S_I)^{\ \ \ A\alpha}_{B\beta}a\]
for all $a\in A$, which means $(S_I)^{\ \ \ A\alpha}_{B\beta}$ are in the centre. Similarly, $(\gamma^I)^{\ \ \ A\alpha}_{B\beta},J^{\ \ \ A\alpha}_{B\beta},\kappa^{\ \ \ A\alpha}_{B\beta}$ are in the centre due to
\[a (\omega^I\la e^{A\alpha})=a\la (\omega^I\tens e^{A\alpha})=\la (a\omega^I\tens e^{A\alpha})=\la (\omega^I\tens e^{A\alpha}a)=\la (\omega^I\tens e^{A\alpha})a=(\omega^I\la e^{A\alpha})a,\]
\[\CJ (e^{A\alpha})a^*=\CJ(a e^{A\alpha})=\CJ(e^{A\alpha}a)=a^*\CJ (e^{A\alpha}),\quad a\gamma(e^{A\alpha})=\gamma(ae^{A\alpha})=\gamma(e^{A\alpha}a)=\gamma(e^{A\alpha})a.\]
Since we assumed that $A_f$ has trivial centre, all our coefficients depend only on $M$. 

Next, the last equation of (\ref{1}) together with (\ref{2}),(\ref{3}) and $\gamma^2=\id$ translate to (\ref{tq2}),(\ref{tq3}). Finally by definition $\dirac=\la\circ\nabla_\CS$ for the Dirac operator acting on spinor gives (\ref{dirac3}). \endproof

\begin{proposition} Suppose that $A_f$ has trivial centre and the product has a QRG with metric and QLC of the form 
\begin{align}\label{ggig}
(\omega^I,\omega^J)= g^{IJ},\quad \nabla \omega^I=-\Gamma^I_{JK}\omega^J\tens\omega^K,\quad \sigma(\omega^I\tens\omega^J)=\omega^J\tens\omega^I
\end{align}
where $g^{IJ},\Gamma^I_{JK}\in C^\infty(M)$ as in \cite{LiuMa2} and $ g^{IJ}$ is symmetric. Then the local spectral triple in Proposition~\ref{local} is fully geometrically realised in sense (\ref{covcliff}),(\ref{cliffreln}) if  $\gamma^I,S$ obey
\begin{align}\label{PG}
\{\gamma^{I},\gamma^{J}\}&=2 g^{IJ}\id,\\
\label{sol2}
S_I=\frac{1}{8}[\gamma_J,\del_I \gamma^{J}&+\Gamma^J_{\ I K}\gamma^{K}]+c_I\id,
\end{align}
where $c_I\in \C^\infty(M)$, $\gamma_I\equiv g_{IJ}\gamma^J$.
\end{proposition}
\proof Firstly, using (\ref{ggig}), (\ref{covcliff}) i.e.
\begin{align*}
 (\id\tens\la)\circ\nabla_{\Omega^1\tens \CS}(\omega^I\tens e^{A\alpha})=\nabla_\CS\circ\la(\omega^I\tens e^{A\alpha})
\end{align*}
gives
\begin{align}\label{PQ}
(\del_J \gamma^{I}+\Gamma^I_{\ J K}\gamma^{K})+S_{J}\gamma^{I}-\gamma^{I}S_{J}=0.
\end{align}

In addition (\ref{cliffreln}) requires
\begin{align*}
\omega^I\la(\omega^J\la e^{A\alpha})+\omega^J\la(\omega^I\la e^{A\alpha})=2(\omega^I,\omega^J)e^{A\alpha}
\end{align*}
giving (\ref{PG}). Next, using (\ref{PQ}),(\ref{PG}), we calculate
\begin{align}\label{Sol}
&[\gamma_I,\del_J \gamma^{I}+\Gamma^I_{\ J K}\gamma^{K}]=[\gamma_I,\gamma^{I}S_{J}-S_{J}\gamma^{I}]=2(n+n_f) S_J-2\gamma_I S_{J}\gamma^{I}.
\end{align}
 On the other hand, we suppose $S_I$ can be written as
\begin{align}\label{sol3}
S_I&=c_{IJK}\gamma^J\gamma^K=\frac{1}{2}c_{IJK}[\gamma^J,\gamma^K]+\frac{1}{2}c_{IJK}\{\gamma^J,\gamma^K\}=\frac{1}{2}c_{IJK}[\gamma^J,\gamma^K]+c_I\id,
\end{align}
where $c_{IJK}\in C^{\infty}(M)$ and $c_I=c_{IJK} g^{JK}$. Then we obtain
\begin{align}
\gamma_J S_I \gamma^J&=\frac{1}{2}c_{IMN}\gamma_J [\gamma^M,\gamma^N] \gamma^J+(n+n_f) c_I\id \nonumber\\
&=\frac{n+n_f-4}{2}c_{IMN}[\gamma^M,\gamma^N]+(n+n_f) c_I\id=(n+n_f-4)S_I+4c_I\id, \label{sol1}
\end{align}
where for the first and third equalities we used (\ref{sol3}), and for the second equality we used
\begin{align*}
\gamma_J\gamma^M\gamma^N\gamma^J&=\{\gamma_J,\gamma^M\}\gamma^N\gamma^J-\gamma^M\{\gamma_J,\gamma^N\}\gamma^J+\gamma^M\gamma^N\gamma_J\gamma^J=4 g^{MN}+(n+n_f-4)\gamma^M\gamma^N. 
\end{align*}
Finally, substituting (\ref{sol1}) into (\ref{Sol}), we obtain (\ref{sol2}).
 \endproof

It remains to solve the conditions for Proposition~\ref{local}, which we will do for $A_f$ the fuzzy sphere. 

\subsection{Spectral triple on $A=C^\infty(M)\tens \C_\lambda[S^2]$}

As this paper follows on from \cite{LiuMa3}, we do not recall all of the details of the QRG on the fuzzy sphere $\C_\lambda[S^2]$, but the main features are that the coordinate algebra has generators $y^i$, $i=1,2,3$ with relations  
\[[y^i,y^j]=2\imath\lambda \eps_{ijk}y^k,\quad \sum_i (y^i)^2=(1-\lambda^2)\]
and a certain differential calculus from\cite[Ex.~1.46]{BegMa} with central basis $s^i$ and associated partial derivatives $\del_i$. The reduced fuzzy sphere $c_\lambda[S^2]$ applies for $\lambda=1/(2k+1)$ with $k\ge 0$  an integer or half-integer and has additional relations making it isomorphic to a matrix algebra $M_{2k+1}(\C)$.

Next, the metric on  tensor product algebra $A=C^\infty(M)\tens \C_\lambda[S^2]$ has the form  \cite{LiuMa3}
\[ g=g_{\mu\nu}\extd x^\mu\tens \extd x^\nu+A_{\mu i}(s^i\tens \extd x^\mu+\extd x^\mu\tens s^i)+h_{ij}s^i\tens s^j, \]
\[\tilde g^{\mu\nu}=(\extd x^\mu,\extd x^\nu),\quad \tilde A^{\mu i}=\tilde A^{i\mu}=-(\extd x^\mu, s^i),\quad (s^i,s^j)=\tilde h^{ij},\]
where 
\[\tilde g_{\mu\nu}=g_{\mu\nu}-h_{ij}\tilde A_\mu^{\ i}\tilde A_\nu^{\ j},\quad \tilde h^{ij}=h^{ij}+\tilde g^{\mu\nu}\tilde A_\mu^{\ i}\tilde A_\nu^{\ j},\quad \tilde A_{\mu}^{\ i}=\tilde g_{\mu\nu}\tilde A^{\nu i}=h^{ij}A_{\mu j},\]
and we will focus on the `round metric' case 
\[ h_{ij}=h\delta_{ij},\quad \del_\alpha h=0.\]
The QLCs have the form
\begin{align*}
\nabla \extd x^\mu=-\Gamma^\mu_{\alpha\beta}\extd x^\alpha\tens \extd x^\beta+F^\mu_{\alpha i}(\extd x^\alpha\tens s^i+s^i\tens \extd x^\alpha)+D^\mu_{ij} s^i\tens s^j,\\
\nabla s^k=E^k_{\alpha\beta}\extd x^\alpha\tens \extd x^\beta+B^k_{\alpha i}(\extd x^\alpha\tens s^i+ s^i\tens \extd x^\alpha)+H^k_{ij} s^i\tens s^j,
\end{align*}
and for the round metric case,
 \begin{align*}
&\Gamma^\sigma_{\mu\nu}=\tilde\Gamma^\sigma_{\mu\nu}-\tilde A_{(\mu}^{\ \  i}F^\sigma_{\nu) i},\quad F^\mu_{\alpha i}=\frac{1}{2}h_{ij}\tilde g^{\beta \mu}\tilde F^j_{\beta\alpha},\quad \tilde F^i_{\mu\nu}= \tilde\nabla_{[\mu}\tilde A_{\nu]}^{\ i}-\epsilon^{i}_{\ jk}\tilde A_\mu^{\ j}\tilde A_\nu^{\ k},\\ 
&D^\mu_{ij}=0, \quad E^k_{\alpha\beta}=-\frac{1}{2}\nabla_{(\alpha}\tilde A_{\beta)}^{\ \ k},\quad B^k_{\alpha i}=-\tilde A_{\mu}^{\ k}F^\mu_{\alpha i}-\frac{1}{2}\epsilon^k_{\ ij} \tilde A_{\alpha}^{\ j},\quad H^i_{jk}=-\frac{1}{2}\epsilon^i_{\ jk}.
\end{align*}

Here, $\tilde\Gamma^\sigma_{\mu\nu}$ are the classical Levi-Civita connection coefficients for the effective metric $\tilde g$ and $\nabla_\alpha$ is the covariant derivative defined by $\Gamma$. We also use the convention $A_{(a}B_{b)}=A_{a}B_{b}+A_{b}B_{a},A_{[a}B_{b]}=A_{a}B_{b}-A_{b}B_{a}$ for any tensors $A,B$ and indices $a,b$.

For later convenience, we note following useful relations
\begin{align}\label{ts0}
\gamma_\mu=\tilde g_{\mu\nu}\tilde\gamma^\nu+\tilde A_{\mu}^{\ i}\gamma_i,\quad\gamma^i=\frac{1}{h}\gamma_i-\tilde A_{\mu}^{\  i}\tilde\gamma^\mu,
\end{align}
which can be deduced from  $\tilde\gamma^\nu=g^{\nu I}\gamma_I=\tilde g^{\mu\nu}\gamma_\mu-\tilde A^{\nu i}\gamma_i$ and $\gamma_j=g_{jI}\gamma^I=A_{\mu j}\tilde\gamma^\mu+h\gamma^j$, where $\tilde\gamma^\mu=\gamma^I|_{I=\mu}$.

\begin{lemma} For $M$ even-dimensional and the above specific metric and QLC on $A=C^\infty(M)\tens \C_\lambda[S^2]$, $\tilde\gamma^{\mu}=\gamma^{a} e_{a}^{\ \mu}$ satisfy the second condition of (\ref{gmma}), and 
\begin{align}\label{ts3}
\gamma^{i}=\frac{1}{\sqrt{h}}\kappa_M\tens\sigma_{i}-\tilde A_{\mu}^{\  i}\tilde\gamma^{\mu}
\end{align}
obey (\ref{PG}). Here, the usual  $\tilde\gamma^\mu$ for the metric $\tilde g^{\mu\nu}$ are viewed acting by the identity on $\CS_f$, $\kappa_M$ is from (\ref{JM2}) and $\sigma_{i}$ are Pauli matrices. 
The formula (\ref{sol2}) then becomes
\[S_{\mu}=\frac{1}{8}\left(2(\omega_{ab})_\mu\gamma^{ab}-h\tilde A_{\mu}^{\ i}\tilde F^i_{\rho\sigma}\tilde\gamma^{\rho\sigma}-2\sqrt{h}\tilde F^i_{\mu\nu}\kappa_M\tilde\gamma^\nu\tens\sigma_{i}+2\imath\tilde A_{\mu}^{\ i}\sigma_i\right)+c_\mu\id,\]
\[S_{i}=-\frac{1}{8}\left(h\tilde F^i_{\mu\nu}\tilde\gamma^{\mu\nu}+2\imath\sigma_i\right)+c_i\id,\]
where
$\tilde\gamma^{\rho\sigma}=\frac{1}{2}[\tilde\gamma^\rho,\tilde\gamma^\sigma]$ and  $\sigma_i$ are viewed acting by the identity on $\CS_M$. 
\end{lemma}
\proof Note that if one wants to write everything more explicitly,  
\[ (\kappa_M\tens\sigma_{i})^{\ \ \ A\alpha}_{B\beta}=(\kappa_M)^{\ A}_{B}(\sigma_i)^{\ \alpha}_\beta,\quad (\tilde\gamma^\mu)^{\ \ \ A\alpha}_{B\beta}=(\tilde\gamma^\mu)^{\ A}_B\delta^{\ \alpha}_\beta,\quad (\sigma_i)_{B\beta}{}^{A\alpha}=(\sigma_i)^{\ \alpha}_{\beta}\delta^{\ A}_B\]
when acting on the product spinor space. 
 
For (\ref{PG}), we need $\{\gamma_i,\tilde\gamma^\mu\}=g_{iI}\{\gamma^I,\tilde\gamma^\mu\}=2g_{iI}g^{I\mu}=0$ and
\begin{align}\label{3eqg}
\{\tilde\gamma^\mu,\tilde\gamma^\nu\}=2\tilde g^{\mu\nu}\id,\quad \{\gamma_i,\gamma_j\}=2h\delta_{ij}\id,\quad \gamma_i\tilde\gamma^\mu+\tilde\gamma^\mu\gamma_i=0.
\end{align}
Using vielbeins, the above equations recover the second condition of (\ref{gmma}) and 
\begin{align}\label{ts4}
\gamma_{i}=\sqrt{h}\kappa_M\tens\sigma_{i}. 
\end{align}
Substituting into the second equation of  (\ref{ts0}), we get (\ref{ts3}). Then one can check that $\tilde\gamma^\mu,\gamma_i$ indeed  solve (\ref{PG}).

For $S_\mu, S_i$, using (\ref{ts0}) and substituting  the metric and QLCs into (\ref{sol2})  we obtain
\begin{align*}S_\mu&=\frac{1}{8}\left(\tilde g_{\rho\sigma}[\tilde\gamma^\rho,\tilde\nabla_\mu\tilde\gamma^\sigma]-h\tilde A_{\mu}^{\ i}\tilde F^i_{\rho\sigma}\tilde\gamma^{\rho\sigma}-2\tilde F^i_{\mu\nu}\gamma_i\tilde\gamma^\nu+\frac{1}{h}\epsilon^{ij}_{\ \ k}\tilde A_{\mu}^{\ k}\gamma_i\gamma_j\right)+c_\mu\id,\\
S_i&=-\frac{1}{8}\left(h\tilde F^i_{\mu\nu}\tilde\gamma^{\mu\nu}+\frac{1}{h}\epsilon^{ij}_{\ \ k}\gamma_j\gamma_k\right)+c_i\id,\end{align*}
which are the results as stated after substituting (\ref{ts4}) and using
\[\tilde g_{\rho\sigma}[\tilde\gamma^\rho,\tilde\nabla_\mu\tilde\gamma^\sigma]=\tilde g_{\rho\sigma}[\gamma^a e^{\ \rho}_a,(\tilde\nabla_\mu e^{\ \sigma}_b)\gamma^b]=\tilde g_{\rho\sigma}e^{\ \rho}_a(\tilde\nabla_\mu e^{\ \sigma}_b)[\gamma^a,\gamma^b]=2(\omega_{ab})_\mu\gamma^{ab}.\]
\endproof

 For integration on $A=C^\infty(M)\tens \C_\lambda[S^2]$, we use the Riemannian integration with measure $\sqrt{-|\tilde g|}$ on $M$ and we use the standard  `integration' on the fuzzy sphere given by the spin 0 component in the orbital  angular momentum expansion as used in \cite{ArgMa1,LirMa2,LiuMa3}. 

\begin{proposition}\label{dirac}We define an inner product on spinors $\Psi_{A\alpha}e^{A\alpha},\Phi_{B\beta}e^{B\beta}$ on the product $A=C^\infty(M)\tens \C_\lambda[S^2]$ by
\[\left\langle \Psi_{A\alpha}e^{A\alpha},\Phi_{B\beta}e^{B\beta}\right\rangle=\int \overline\Psi\Phi,\]
where $\overline{\Psi}=-\imath\Psi^{\dagger}\gamma^0$ with $\gamma_0$ just acting on the $\CS_M$ indices. Then the above construction results in a real spectral triple at the pre-Hilbert space level with Dirac operator 
\begin{align*}
\dirac (\Psi_{A\alpha} e^{A\alpha})&=\Big(\Big(\tilde\dirac_M-\tilde A_{\mu}^{\  i}\tilde\gamma^{\mu}\tens\Big(\del_i-\frac{\imath}{2}\sigma_i\Big)+\frac{\sqrt{h}}{8}\tilde F_{\mu\nu}^{i}\tilde\gamma^{\mu\nu}\kappa_M\tens\sigma_i+\kappa_M\tens\dirac_f\Big)\Psi\Big)_{A\alpha} e^{A\alpha},
\end{align*}
where
\begin{align}\label{df}
\dirac_f=\frac{1}{\sqrt{h}}\Big(\sigma_{i}\del_i-\frac{3\imath}{4}\Big).
\end{align}
We also take
\begin{align}\label{Jq}
J=q J_M\tens \sigma_2,\quad |q|^2=1,\quad q\in\C,\quad \epsilon=-1,\quad \epsilon'=1.
\end{align} 
\end{proposition}
\proof (1) We first show that $\CJ,\eps,\eps'$ as stated together with $S_\mu,S_i$ from the Lemma and the product Clifford structure as above obey the conditions for a spectral triple at the local tensorial level, which means that they need to satisfy (\ref{tq2})(they don't need to satisfy (\ref{tq3}) because we consider the total dimension to be odd).

 At the matrix level, we suppose $J^{\ \ \ A\alpha}_{B\beta}=J^{\ A}_{B}J^{\ \alpha}_{\beta}$ or $J=J_M\tens J_f$. Then the third equation of (\ref{tq2}) reduces to
\[J_f \sigma_i^*=-\epsilon'\sigma_i J_f,\]
 which is only solvable in the case $\epsilon'=1$ and the solution is
\[J_f=q\sigma_2,\quad q\in C^\infty(M).\]
This gives $J$ of the form stated. Moreover, the first equation of (\ref{tq2}) gives
\[|q|^2=1,\quad \epsilon=-1.\]
Now to solve the second equation of (\ref{tq2}), we require
\[ S_\mu \sigma_2=\sigma_2 S_\mu^*,\quad  S_i \sigma_2= \sigma_2 S_i^*, \quad q\in\C.\]
Substituting $S_\mu,S_i$, one can check the above first two conditions hold only in the case $c_I=c_I^*$.

(2) By using the expression of $S_\mu,S_i$ in the above lemma, we calculate
\begin{align*}
\gamma^I S_I=&\frac{1}{8}\Big(2(\omega_{ab})_\mu\tilde\gamma^\mu\gamma^{ab}+\sqrt{h}\tilde F_{\mu\nu}^{i}\tilde\gamma^{\mu\nu}\kappa_M\tens\sigma_i-\frac{6\imath}{\sqrt{h}}\kappa_M+4\imath\tilde A_{\mu}^{\  i} \tilde\gamma^\mu\tens\sigma_i\Big)+c_I\gamma^I. 
\end{align*}
Then substituting $\gamma^I S_I$ and (\ref{ts3}) into (\ref{dirac3}), we get
\begin{align*}
\dirac (\Psi_{A\alpha} e^{A\alpha})&=\Big(\Big(\tilde\dirac_M-\tilde A_{\mu}^{\  i}\tilde\gamma^{\mu}\tens\Big(\del_i-\frac{\imath}{2}\sigma_i\Big)+\frac{\sqrt{h}}{8}\tilde F_{\mu\nu}^{i}\tilde\gamma^{\mu\nu}\kappa_M\tens\sigma_i+\kappa_M\tens\dirac_f+c_I\gamma^I\Big)\Psi\Big)_{A\alpha} e^{A\alpha}. 
\end{align*}

(3) Next, we need $\dirac$ to be hermitian. We calculate
\begin{align*}
&\langle \dirac (\Psi_{A\alpha}e^{A\alpha}),\Phi_{B\beta}e^{B\beta}\rangle\\
=&-\imath\int\Big(\Big(\tilde\dirac_M-\tilde A_{\mu}^{\  i}\tilde\gamma^{\mu}\tens\Big(\del_i-\frac{\imath}{2}\sigma_i\Big)+\frac{\sqrt{h}}{8}\tilde F_{\mu\nu}^{i}\tilde\gamma^{\mu\nu}\kappa_M\tens\sigma_i+\kappa_M\tens\dirac_f+c_I\gamma^I\Big)\Psi\Big)^\dagger\gamma^0\Phi\\
=&\int \overline\Psi\tilde\dirac_M\Phi+\imath\int \Psi^\dagger\gamma^0\Big(\tilde A_{\mu}^{\  i}\tilde\gamma^{\mu}\tens\Big(\del_i-\frac{\imath}{2}\sigma_i\Big)-\frac{\sqrt{h}}{8}\tilde F_{\mu\nu}^{i}\tilde\gamma^{\mu\nu}\kappa_M\tens\sigma_i-\kappa_M\tens\dirac_f+c_I^*\gamma^I\Big)\Phi\\
=&\int \overline\Psi\Big(\tilde\dirac_M-\tilde A_{\mu}^{\  i}\tilde\gamma^{\mu}\tens\Big(\del_i-\frac{\imath}{2}\sigma_i\Big)+\frac{\sqrt{h}}{8}\tilde F_{\mu\nu}^{i}\tilde\gamma^{\mu\nu}\kappa_M\tens\sigma_i+\kappa_M\tens\dirac_f-c_I^*\gamma^I\Big)\Phi\\
=&\langle \Psi_{A\alpha}e^{A\alpha},\dirac(\Phi_{B\beta}e^{B\beta})\rangle
\end{align*}
provided $c_I=-c_I^*$. For the second equality, we used the hermitian property of $\dirac_M$, hermitian conditions (\ref{ts5})(\ref{ts2}) and dropped total derivatives  $\del_i$. Since the  $c_I$ are real, we are forced to take $c_I=0$.

Next, we can show $\CJ$ is antilinear and an isometry or anti-isometry 
\begin{align*}
&\<\CJ(\Phi_{A\alpha}e^{A\alpha}),\CJ(\Psi_{B\beta}e^{B\beta})\>=\<J^{\ \ \ A\alpha}_{C\gamma}\Phi_{A\alpha}^*e^{C\gamma},J^{\ \ \ B\beta}_{D\delta}\Psi_{B\beta}^*e^{D\delta}\>\\
&=\int \overline{J\Phi^*}J\Psi^*=-\imath\int (J\Phi^*)^\dagger\gamma^0 J\Psi^*=-\imath\eps'_M\int \Phi^{\intercal} J^\dagger\gamma^0 J\Psi^*=-\imath\eps'_M\int \Phi^{\intercal} \gamma^{0*}\Psi^*\\
&=-\imath\eps'_M\int (\Phi^{\intercal} \gamma^{0*}\Psi^*)^{\intercal}=\imath\eps'_M\int \Psi^\dagger\gamma^{0}\Phi=-\eps'_M\int \overline\Psi\Phi=-\eps'_M\<\Psi_{A\alpha}e^{A\alpha},\Phi_{B\beta}e^{B\beta}\>
\end{align*}
where for the fifth equality we used $J^\dagger\gamma^0 J=(J_M^\dagger\tens\sigma_2)\gamma^0 (J_M\tens\sigma_2)=J_M^\dagger\gamma^0 J_M=\eps'_M\gamma^{0*}$.
\endproof

This solves the requirements at formal level where we did not yet complete the spinor space to a Hilbert space. One can take $\eps'_M=-1$ to obtain what formally looks like an antilinear isometry for $\CJ$, and one can also check that $\<\Psi,\Phi\>^*=\<\Phi,\Psi\>$. However, we do not have positivity and hence we do not have a Hilbert space as such.

\section{Spacetime spinor multiplets from the model}\label{sec:multi}

Next, to analyse the particle content from the point of view of spacetime, we need to recall the spinor geometry and Dirac operator $\imath\sqrt{h}\dirac_f$ introduced in \cite{LirMa2} which we saw emerges in (\ref{df}). From that formula and $\epsilon^{ij}_{\ \ k}\del_i\del_j=\del_k$, we have
\begin{equation}\label{dsquare} -(\imath\sqrt{h}\dirac_f)^2=\del_i^2-\frac{1}{2}(\imath\sqrt{h}\dirac_f)-\frac{3}{16}. \end{equation}
We let $A_l\subset A_f$ be the subspace the coordinate algebra of the fuzzy sphere of orbital angular momentum $l$ (so that $\del_i^2=-l(l+1)$). Then $A_l\tens \C^2\subset \CS_f$, where $\C^2$ is the spin 1/2 representation of the spinor indices, decomposes as
\[ A_l\tens \C^2= S_l^+\oplus \CS_l^-,\quad \imath\sqrt{h}\dirac_f|_{\CS_l^\pm}={1\over 4}\pm (l+{1\over 2}) \]
according the roots of (\ref{dsquare}) for the eigenvalues of $\imath\sqrt{h}\dirac_f$ as explained in \cite{LirMa2}. Here  $\CS_l^\pm$ are the corresponding eigenspaces except that there turns out to be no subspace $\CS_0^{-}$ if $l=0$.   In fact, these eigenspaces are each irreps for the total angular momentum on the fuzzy sphere  
\[ J_i=L_i+ {\sigma_i\over 2}=\imath(\del_i - {\imath\over 2}\sigma_i)   \]
with total spin $j=l\pm {1\over 2}$ on $S_l^\pm$. This follows from the formula
\begin{equation}\label{Jsquare} (\del_i-\frac{\imath}{2}\sigma_i)^2=\del_i^2-\imath\sqrt{h}\dirac_f,\end{equation}
where the left hand side is  $-j(j+1)$ if the total spin is $j$, while $\del_i^2=-l(l+1)$ on the whole of $A_l\tens\C^2$, which for $j\ge 0$,  uniquely determines $j=l\pm{1\over 2}$ for the two eigenspaces when $l>0$ and $j={1\over 2}$ when $l=0$. One can check that $S_l^\pm$ is indeed $2j+1$-dimensional for these values of $j$ and hence is an irrep of $SU(2)$ with this Yang-Mills `charge'.  In the case of the reduced fuzzy sphere where the deformation parameter is $\lambda=1/(2k+1)$, we only sum  $0\le l\le 2k$ .  For example,
\[ k={1\over 2}:\quad \CS_f=M_2\tens \C^2= \CS_0^+\oplus \CS_1^-\oplus \CS_1^+= 2\oplus 2\oplus 4,\]
\[ k=1;\quad \CS_f=M_3\tens \C^2= \CS_0^+\oplus \CS_1^-\oplus \CS_1^+\oplus \CS_2^-\oplus \CS_2^+= 2\oplus 2\oplus 4 \oplus 4\oplus 6,\]
where on the right we put the corresponding $2j+1$-dimensional irreps of $SU(2)$.  The explicit decomposition of $A_l\tens \C^2$ into $S_l^\pm$ is provided by the relevant Clebsch-Gordan coefficients, but we do not need their explicit form to proceed.

We now decompose any spinor $\Psi$ on the total space as
\[ \Psi=\sum_{l\ge 0, \pm} \Psi_{l,\pm},\quad  \Psi_{l,\pm}\in \CS_M\tens \CS_l^\pm\]
except that for $l=0$ we only have $\Psi_{0,+}$. Here each $\Psi_{l,\pm}$ appears as a multiplet of spinors on $M$ forming the $j=l\pm{1\over 2}$ representation of $SU(2)$ now viewed as an internal symmetry. We mostly do not need to explicitly describe the allowed fields according to $ \CS_l^\pm\subset A_l\tens \C^2$, but viewed in the latter space they have the form
\begin{align}
\Psi_{l,\pm}&=(\psi_{l,\pm})_{i_1...i_l A\alpha}y^{i_1}y^{i_2}...y^{i_l}e^{A\alpha}
\end{align}
for {\em certain} totally  symmetric and traceless tensors in the $y^i$ fuzzy sphere coordinates indices and two further  indices for spinors on $M$ and internally on the fuzzy sphere. These are constrained so that $\Psi_{l,\pm}$ lives in $S_l^\pm$ with regard to its fuzzy spinor indices.
 The orbital angular momentum generator $\del_i$ is given on these by application of the totally antisymmetric tensor $\eps_{i\cdot\cdot}$ as a matrix acting in turn on the tensor indices as for scalar fields in \cite{ArgMa1,LiuMa3}. The $\sigma_i$ just acts on the $\C^2$ component and  their combination defines $J_i$ which acts as 
\[ J_i\Psi_{l,\pm}= T^i_j\Psi_{l,\pm},\]
where $T^i_j$ is the action of $J_i$ in the spin $j=l\pm{1\over 2}$ representation of $SU(2)$. Explicitly, putting in all the tensor and spinor indices,
\begin{equation}\label{expT} (T^i_j\psi_{l,\pm})_{i_1\cdots i_l A\alpha}=-\imath\sum_p \eps_{ii_p k} \psi_{l,\pm}{}_{i_1\cdots k\cdots i_l A\alpha}+ {1\over 2}(\sigma_i\psi_{l,\pm})_{i_1\cdots i_l A\alpha},
 \end{equation}
\begin{equation}\label{expsigma}  (\sigma_i\psi_{l,\pm})_{i_1\cdots i_l A\alpha}= (\sigma_i)_\alpha{}^\beta\psi_{l,\pm}{}_{i_1\cdots i_l A\beta}.\end{equation}
Either way, Proposition~\ref{dirac} tells us
\begin{align*}
\dirac (\Psi_{A\alpha} e^{A\alpha})=\sum_{l\ge 0,\pm}\Big(\Big(\tilde\dirac_{A,j} -\imath m_{l,\pm}\kappa_M+\frac{\sqrt{h}}{8}\tilde F_{\mu\nu}^i\tilde\gamma^{\mu\nu}\kappa_M\tens\sigma_i\Big)\psi_{l,\pm}\Big)_{ i_1...i_lA\alpha}y^{i_1}y^{i_2}...y^{i_l} e^{A\alpha},
\end{align*}
where 
\[ \tilde\dirac_{A,j}=\tilde\dirac_M+\imath  T^i_j\tilde A_{\mu}^{\ i}\tilde\gamma^\mu,\quad m_{l,\pm}={1\over\sqrt{h}} \Big( {1\over 4}\pm \Big(l+{1\over 2}\Big) \Big), \]
 is the minimally-coupled Dirac operator for a spinor field coupled to an $SU(2)$ Yang-Mills field of `spin' $j=l\pm{1\over 2}$ (which now appears as the Yang-Mills charge from the point of view of spacetime) and mass as shown. In component terms 
 \begin{equation} (\dirac \psi_{l,\pm})_{i_1... i_l A\alpha}= \Big(\Big(\tilde\dirac_{A,j} -\imath m_{l,\pm}\kappa_M+\frac{\sqrt{h}}{8}\tilde F_{\mu\nu}^i\tilde\gamma^{\mu\nu}\kappa_M\tens\sigma_i\Big)\psi_{l,\pm}\Big)_{ i_1...i_l A\alpha},\end{equation}
 which can be explicitly computed using (\ref{expT})-(\ref{expsigma}). We see that there is an induced `chiral mass' $m_{l,\pm}$ (or mass shift if the field on the product has such a mass) and there is additionally a coupling via $\sigma_i$ to the Yang-Mills curvature. 

Next, we consider the Dirac operator with a positive mass acting on $\psi_{l,\pm}$
\begin{align*}
((\dirac-m) \psi_{l,\pm})_{i_1... i_l A\alpha}= \Big(\Big(\tilde\dirac_{A,j} -(m+\imath m_{l,\pm}\kappa_M)+\frac{\sqrt{h}}{8}\tilde F_{\mu\nu}^i\tilde\gamma^{\mu\nu}\kappa_M\tens\sigma_i\Big)\psi_{l,\pm}\Big)_{ i_1...i_l A\alpha}.
\end{align*}
To extract $\kappa_M$ in the mass term, we apply a chiral transformation \cite{freedman}
\[\tilde\psi_{l,\pm}=e^{-\imath \alpha\kappa_M}\psi_{l,\pm},\]
where $\alpha=\alpha_{l,\pm}$ is constant, then by using 
\[e^{\imath \alpha\kappa_M}\tilde\gamma^\mu=\tilde\gamma^\mu e^{-\imath \alpha\kappa_M},\quad e^{-\imath \alpha\kappa_M}\kappa_M=\kappa_M e^{-\imath \alpha\kappa_M},\quad e^{2\imath\alpha \kappa_M}=\cos 2\alpha+\imath\kappa_M \sin 2\alpha,\]
we obtain 
\begin{align*}
((&\dirac-m) \psi_{l,\pm})_{i_1... i_l A\alpha}\\
&=\Big(e^{-\imath \alpha\kappa_M}\Big(\tilde\dirac_{A,j} -e^{2\imath \alpha\kappa_M}(m+\imath m_{l,\pm}\kappa_M)-\frac{\imath \sqrt{h}}{8}\tilde F_{\mu\nu}^i\tilde\gamma^{\mu\nu}e^{2\imath (\alpha+\frac{\pi}{4})\kappa_M}\tens\sigma_i\Big)\tilde\psi_{l,\pm}\Big)_{ i_1...i_l A\alpha}. 
\end{align*}
For $m\ne 0$, if we set 
\begin{align}\label{angle}
\tan 2\alpha=-\frac{m_{l,\pm}}{m},\quad -\frac{\pi}{4}<\alpha<\frac{\pi}{4}
\end{align}
then above mass term becomes
\[m_{l,\pm}'\equiv e^{2\imath \alpha\kappa_M}(m+\imath m_{l,\pm}\kappa_M)=(\cos 2\alpha+\imath\kappa_M \sin 2\alpha)(m+\imath m_{l,\pm}\kappa_M)=\frac{m}{\cos 2\alpha}=\sqrt{m^2+m^2_{l,\pm}}\]
which is positive. If $m=0$ we have instead
\begin{align}\label{angle2}
\alpha= -{\rm sign}( m_{l,\pm}){\pi\over 4}, \quad m'_{l,\pm}=|m_{l,\pm}| 
\end{align}
by solving separately. Then 
\begin{align*}
&((\dirac-m) \psi_{l,\pm})_{i_1... i_l A\alpha}=\Big(e^{-\imath \alpha\kappa_M}\Big(\tilde\dirac_{A,j} -m_{l,\pm}'-\frac{\imath \sqrt{h}}{8}\tilde F_{\mu\nu}^i\tilde\gamma^{\mu\nu}e^{2\imath (\alpha+\frac{\pi}{4})\kappa_M}\tens\sigma_i\Big)\tilde\psi_{l,\pm}\Big)_{ i_1...i_l A\alpha}.
\end{align*}

Now we can write down the Dirac action for $\Psi_{A\alpha}e^{A\alpha}$ on the product,
\begin{align*}
S&=\<\Psi_{A\alpha}e^{A\alpha},((\dirac-m)\Psi)_{B\beta}e^{B\beta}\>=\int\overline\Psi(\dirac-m)\Psi\\
&=\sum_{l'\ge 0,\pm'}\sum_{l\ge 0,\pm}\int(\overline{\tilde\psi}_{l',\pm’})_{ i_1'...i_{l'}'}\Big(\Big(\tilde\dirac_{A,j}- m_{l,\pm}'-\frac{\imath \sqrt{h}}{8}\tilde F_{\mu\nu}^i\tilde\gamma^{\mu\nu}e^{2\imath (\alpha+\frac{\pi}{4})\kappa_M}\tens\sigma_i\Big)\tilde\psi_{l,\pm}\Big)_{ i_1...i_{l}}\\
&\qquad\qquad\qquad\qquad y^{i_1'}...y^{i_{l'}'}y^{i_1}...y^{i_l}\\
&=\sum_{l\ge 0,\pm}\int_M c_{l}\overline{\tilde\psi}_{l,\pm}\Big(\tilde\dirac_{A,j}- m_{l,\pm}'-\frac{\imath \sqrt{h}}{8}\tilde F_{\mu\nu}^i\tilde\gamma^{\mu\nu}e^{2\imath (\alpha+\frac{\pi}{4})\kappa_M}\tens\sigma_i\Big)\tilde\psi_{l,\pm},
\end{align*}
where we suppressed the multiplet indices. For the last equality, we used as in \cite{LiuMa3}
\[\int_{\C_{\lambda}[S^2]} A_{i_1...i_l} B_{ i_1'...i_{l'}'} y^{i_1}y^{i_2}...y^{i_l}y^{i_1'}y^{i_2'}...y^{i_{l'}'}=c_l A_{i_1...i_l} B_{ i_1...i_{l'}}\delta_{l,l'},\]
where $A,B$ are  totally symmetric and traceless and $c_l$ are some real counting factors and we also used the orthogonality of $\tilde\psi_{l,+},\tilde\psi_{l,-}$ for the inner product on $\CS_f$ since they are living in different eigenspaces of $\imath\sqrt{h}\dirac_f$. Furthermore, we can absorb $c_l$ into $\tilde\psi_{l,\pm}$ which gives
\begin{align*}
S=\sum_{l\ge 0,\pm}\int_M\overline{\tilde\psi}_{l,\pm}\Big(\tilde\dirac_{A,j}- m_{l,\pm}'-\frac{\imath \sqrt{h}}{8}\tilde F_{\mu\nu}^i\tilde\gamma^{\mu\nu}e^{2\imath (\alpha+\frac{\pi}{4})\kappa_M}\tens\sigma_i\Big)\tilde\psi_{l,\pm}.
\end{align*}

To calculate the last term in this action, we expand $\tilde\psi_{l,\pm}$ by
\[\tilde\psi_{l,\pm}=\sum_{j_3}\tilde\psi_{l,\pm,j_3},\] 
where $\tilde\psi_{l,\pm,j_3}$ are orthogonal eigenstates of $T^3_j$ with eigenvalue $j_3$ where $-j\le j_3\le j$ for $j=l\pm{1\over 2}$. For $A_l\tens \C^2$ we choose $\C^2$ an orthogonal eigenbasis of $\sigma_3/2$ (this specifies the index $
\alpha$ above). Then using the Clebsch-Gordan coefficients we can write the components $\tilde\psi'$ with respect to this basis of $\C^2$ as 
\begin{align}\label{cg1}
\tilde\psi_{l,\pm,j_3}=\pm\sqrt{{l\pm j_3+{1\over 2}\over 2l+1}}\tilde\psi_{l,j_3-{1\over 2},{1\over 2}}'+\sqrt{{l\mp j_3+{1\over 2}\over 2l+1}}\tilde\psi_{l,j_3+{1\over 2},-{1\over 2}}'
\end{align}
for $l\ge 1$.  The first and second terms  on the right hand side  are the common eigenstates of $L^2_i,L_i,{1\over 2}\sigma_3$ with eigenvalues $l(l+1),j_3\mp{1\over 2},\pm{1\over 2}$ respectively. Conversely, we have
\begin{align}\label{cg2}
\tilde\psi_{l, j_3\mp{1\over 2},\pm{1\over 2}}'=\sqrt{{l\pm j_3+{1\over 2}\over 2l+1}}\tilde\psi_{l,+,j_3}\mp \sqrt{{l\mp j_3+{1\over 2}\over 2l+1}}\tilde\psi_{l,-,j_3}. 
\end{align}
Then we can calculate
\begin{align*}
\sum_{l\ge 1,\pm}&\int_M\overline{\tilde\psi}_{l,\pm}(\tilde F_{\mu\nu}^{i}\tilde\gamma^{\mu\nu}e^{2\imath (\alpha+\frac{\pi}{4})\kappa_M}\tens\sigma_i)\tilde\psi_{l,\pm}\\
=&\sum_{l\ge 1,\pm,j_3}\int_M\overline{\tilde\psi}_{l,\pm,j_3}(\tilde F_{\mu\nu}^{i}\tilde\gamma^{\mu\nu}e^{2\imath (\alpha+\frac{\pi}{4})\kappa_M}\tens\sigma_i)\tilde\psi_{l,\pm,j_3}\\
=&\sum_{l\ge 1,\pm,j_3}\int_M {l\pm j_3+{1\over 2}\over 2l+1}\overline{\tilde\psi'}_{l,j_3-{1\over 2},{1\over 2}}(\tilde F_{\mu\nu}^{i}\tilde\gamma^{\mu\nu}e^{2\imath (\alpha+\frac{\pi}{4})\kappa_M}\tens\sigma_i)\tilde\psi_{l,j_3-{1\over 2},{1\over 2}}'\\
&\quad\quad\quad +{l\mp j_3+{1\over 2}\over 2l+1}\overline{\tilde\psi'}_{l,j_3+{1\over 2},-{1\over 2}}(\tilde F_{\mu\nu}^{i}\tilde\gamma^{\mu\nu}e^{2\imath (\alpha+\frac{\pi}{4})\kappa_M}\tens\sigma_i)\tilde\psi_{l,j_3+{1\over 2},-{1\over 2}}'\\
=&\sum_{l\ge 1,\pm,j_3}\int_M \tilde F_{\mu\nu}^{3}\Big( {l\pm j_3+{1\over 2}\over 2l+1}\overline{\tilde\psi'}_{l,j_3-{1\over 2},{1\over 2}}\tilde\gamma^{\mu\nu}e^{2\imath (\alpha+\frac{\pi}{4})\kappa_M}\tilde\psi_{l,j_3-{1\over 2},{1\over 2}}'\\
&\quad\quad\quad-{l\mp j_3+{1\over 2}\over 2l+1}\overline{\tilde\psi'}_{l,j_3+{1\over 2},-{1\over 2}}\tilde\gamma^{\mu\nu}e^{2\imath (\alpha+\frac{\pi}{4})\kappa_M}\tilde\psi_{l,j_3+{1\over 2},-{1\over 2}}'\Big)\\
=&\sum_{l\ge 1,\pm,j_3}\frac{1}{2l+1}\int_M \tilde F_{\mu\nu}^{3}\Big((j_3\pm j_3)\overline{\tilde\psi}_{l,+,j_3}\tilde\gamma^{\mu\nu}e^{2\imath (\alpha+\frac{\pi}{4})\kappa_M}\tilde\psi_{l,+,j_3}\\
&\quad\quad\quad\quad\quad +(-j_3\pm j_3)\overline{\tilde\psi}_{l,-,j_3}\tilde\gamma^{\mu\nu}e^{2\imath (\alpha+\frac{\pi}{4})\kappa_M}\tilde\psi_{l,-,j_3}\Big)\\
=&\sum_{l\ge 1,j_3}\frac{2}{2l+1}\int_M \tilde F_{\mu\nu}^{3}j_3\left(\overline{\tilde\psi}_{l,+,j_3}\tilde\gamma^{\mu\nu}e^{2\imath (\alpha+\frac{\pi}{4})\kappa_M}\tilde\psi_{l,+,j_3}-\overline{\tilde\psi}_{l,-,j_3}\tilde\gamma^{\mu\nu}e^{2\imath (\alpha+\frac{\pi}{4})\kappa_M}\tilde\psi_{l,-,j_3}\right)\\
=&\sum_{l\ge 1,\pm,j_3}\frac{\pm 2}{2l+1}\int_M \tilde F_{\mu\nu}^{3}j_3\overline{\tilde\psi}_{l,\pm,j_3}\tilde\gamma^{\mu\nu}e^{2\imath (\alpha+\frac{\pi}{4})\kappa_M}\tilde\psi_{l,\pm,j_3}\\
=&\sum_{l\ge 1,\pm,j_3}\frac{\pm 2}{2l+1}\int_M \overline{\tilde\psi}_{l,\pm,j_3}\tilde\gamma^{\mu\nu}e^{2\imath (\alpha+\frac{\pi}{4})\kappa_M}\tilde F_{\mu\nu}^{3}T^3_j\tilde\psi_{l,\pm,j_3}\\
=&\sum_{l\ge 1,\pm}\frac{\pm 2}{2l+1}\int_M \overline{\tilde\psi}_{l,\pm}\tilde\gamma^{\mu\nu}e^{2\imath (\alpha+\frac{\pi}{4})\kappa_M}\tilde F_{\mu\nu}^{a}T^a_j\tilde\psi_{l,\pm},
\end{align*}
where we used orthogonality of $\tilde\psi_{l,\pm,j_3}$ for the first equality, (\ref{cg1}) and orthogonality of $\tilde\psi'$ for the second equality. For the third equality, we notice $\sigma_1,\sigma_2$ change the eigenstates $\tilde\psi_{l,l_3,\pm{1\over 2}}'$ into $\tilde\psi_{l,l_3,\mp{1\over 2}}'$ and $\sigma_3\tilde\psi_{l,l_3,\pm{1\over 2}}'=\pm\tilde\psi_{l,l_3,\pm{1\over 2}}'$ then use orthogonality of $\tilde\psi'$. In addition, we used (\ref{cg2}) for the fourth equality and the eigenvalue of $T^3_j$ for the equality before the last. For the last equality, we used that $T^{1,2}_j$ can written in terms of raising and lowering operators with result which are then orthogonal to the $j_3$ eigenspace. 

 For the $l=0$ case, we have more directly, 
\[\int_M\overline{\tilde\psi}_{0,+}(\tilde F_{\mu\nu}^{i}\tilde\gamma^{\mu\nu}e^{2\imath (\alpha+\frac{\pi}{4})\kappa_M}\tens\sigma_i)\tilde\psi_{0,+}=2\int_M\overline{\tilde\psi}_{0,+}(\tilde F_{\mu\nu}^{i}\tilde\gamma^{\mu\nu}e^{2\imath (\alpha+\frac{\pi}{4})\kappa_M}T^i_{\frac{1}{2}})\tilde\psi_{0,+}.\]

As a result the final action is
\begin{align}\label{S}
S=&\sum_{l\ge 0,\pm}\int_M\overline{\tilde\psi}_{l,\pm}\left(\tilde\dirac_{A,j}- m_{l,\pm}'\pm\frac{\imath\sqrt{h}}{4(2l+1)}\tilde F_{\mu\nu}^{a}T^a_j\tilde\gamma^{\mu\nu}e^{2\imath (\alpha+\frac{\pi}{4})\kappa_M}\right)\tilde\psi_{l,\pm},
\end{align}
where again the summation is understood that when $l=0$ one only has the $+$ case and $\alpha$ is given by (\ref{angle})(\ref{angle2}) and the mass term $m_{l,\pm}'$ is positive and satisfies
\begin{align}
m_{l,\pm}'^2=m^2+{1\over h} \Big( l+{1\over 2}\pm{1\over 4}  \Big)^2.
\end{align}

As noted, both the noncommutative differential geometry and the action of the symmetry $SU_2$ used in the  analysis above apply also for reduced fuzzy spheres where $\lambda=1/(2k+1)$ and now with the restriction $0\le l\le 2k$. Note that this formula for the masses still makes sense for $k=l=0$ as an extreme case even though in this case  $c_1[S^2]=\C$ and we do not then have the full picture of a nontrivial quantum Riemannian geometry. In this case $S_f=\C\tens\C^2$ and a massless fermion singlet becomes a massive doublet.

\section{Concluding remarks and interpretation}\label{sec:con}

We recall that the physical covariant derivative and physical field strength are
\[\tilde\dirac_{\bar A,j}=\tilde\dirac_M-\imath g  T^i_j\bar A_{\mu}^{\ i}\tilde\gamma^\mu,\quad \bar F^i_{\mu\nu}=\tilde\nabla_{[\mu}\bar A_{\nu]}^{\ i}+g\epsilon^i{}_{jk}\bar A_{\mu}^{\ j}\bar A_{\nu}^{\ k}\]
and if we think of the internal $SU(2)$ as $SU(2)_w$ for the electroweak case then, as in  \cite{LiuMa3}, 
\[g=\sqrt{\frac{16\pi G}{h}}\approx 0.64.\] 
Then written in terms of physical quantities $\bar A,\bar F$, the action is
\begin{align}\label{S1}
S=&\sum_{l\ge 0,\pm}\int_M\overline{\tilde\psi}_{l,\pm}\left(\tilde\dirac_{\bar A,j}- m_{l,\pm}'\mp\frac{\imath \sqrt{\pi G}}{2l+1}\bar F_{\mu\nu}^{a}T^a_j\tilde\gamma^{\mu\nu}e^{2\imath (\alpha+\frac{\pi}{4})\kappa_M}\right)\tilde\psi_{l,\pm},
\end{align}
where the mass is positive and satisfies
\begin{align}\label{mlp}
m_{l,\pm}'^2=m^2+{g^2\over 16\pi} \Big( l+{1\over 2}\pm{1\over 4} \Big)^2 m^2_p\approx m^2+  0.008\Big( l+{1\over 2}\pm{1\over 4} \Big)^2m_p^2.
\end{align} 
Here, $m_p=\frac{1}{\sqrt{G}}$ is Planck mass. 

This is an infinite tower as we sum over all $l$, but we can get a finite multiplet if we take the fibre to be a  reduced fuzzy sphere. For example, for  $c_\frac{1}{2}[S^2]\cong M_2(\C)$ case, the action is
\begin{align*}
S=&\int_M\overline{\tilde\psi}_{0,+}\left(\tilde\dirac_{\bar A,\frac{1}{2}}- m_{0,+}'-\imath \sqrt{\pi G}\bar F_{\mu\nu}^{a}T^a_\frac{1}{2}\tilde\gamma^{\mu\nu}e^{2\imath (\alpha_{0,+}+\frac{\pi}{4})\kappa_M}\right)\tilde\psi_{0,+}\\
&+\overline{\tilde\psi}_{1,+}\left(\tilde\dirac_{\bar A,\frac{3}{2}}- m_{1,+}'-\frac{\imath\sqrt{\pi G}}{3}\bar F_{\mu\nu}^{a}T^a_\frac{3}{2}\tilde\gamma^{\mu\nu}e^{2\imath (\alpha_{1,+}+\frac{\pi}{4})\kappa_M}\right)\tilde\psi_{1,+}\\
&+\overline{\tilde\psi}_{1,-}\left(\tilde\dirac_{\bar A,\frac{1}{2}}- m_{1,-}'+\frac{\imath \sqrt{\pi G}}{3} \bar F_{\mu\nu}^{a}T^a_\frac{1}{2}\tilde\gamma^{\mu\nu}e^{2\imath (\alpha_{1,-}+\frac{\pi}{4})\kappa_M}\right)\tilde\psi_{1,-},
\end{align*}
where $\tilde\psi_0$ is a Yang-Mills doublet (i.e. $j={1\over 2}$) of spinor fields in $M$, $\tilde\psi_{1,+}$ is a Yang-Mills quadruplet $(j=3/2$) and $\tilde\psi_-$ is another Yang-Mills doublet, with varying masses from (\ref{mlp}). Since these masses are already Planckian, the  curvature will be relatively suppressed for physical values where $||\bar F||\ll m_p^2$ (taking $l=0$, for example).  Planckian-mass spinors do not feature directly in the Standard Model but could enter, for example, in a see-saw mechanism for neutrino oscillations. This has Planckian right-handed neutrinos\cite{Sch}, albeit these are $SU(2)_w$ scalars which is not our case. 

Another scenario is that the $SU(2)$ is an as yet unobserved extremely weak gauge field possibly related to some approximate symmetry (not necessarily $SU(2)_f$ as discussed in \cite{LiuMa3}), in such a way that the mass corrections are of a Standard Model level. In this case we see a particularly distinctive pattern of spinor matter fields and masses, e.g.  for $k=1/2$ and say $m=0$, we now get induced masses for the above three fields as
\[ m'_{0,+}= {3\over 4\sqrt{h}}, \quad   m'_{1,-}=\frac{5}{3}m'_{0,+},\quad m'_{1,+}=\frac{7}{3}m'_{0,+},\]
where $h$ in (\ref{S}) is  chosen so that the masses are of the Standard Model level. As explained in \cite{LiuMa3}, this  requires the effective  structure constant $\alpha$ to be extremely weak, comparable to gravity for this scale of mass. Also note that in this scenario, the Yang-Mills curvature term is relatively suppressed in (\ref{S}) if 
\[ ||\bar F||\ll (m'_{0,+})^2.\] 
Recall that $\tilde\psi_{0,+}, \tilde\psi_{1,-}$ here are $SU(2)$ doublets and $\tilde\psi_{1,+}$ is an $SU(2)$ quadruplet, and we see that their masses are in ratio $1:5/3:7/3$.

This extends to analysis in \cite{LiuMa3}, where we considered scalar fields on the product, to the more relevant spinor case. We obtained a distinctive signature for the resulting spinor multiplets viewed on $M$, which as far as we know is not observed in the Standard Model, but acts as proof of concept which can be developed further for different choices of noncommutative fibre algebra $A_f$.


\begin{thebibliography}{ggghhh}

\bibitem{ArgMa1} J. Argota-Quiroz and S. Majid, Fuzzy and discrete black hole models, Class. Quantum Grav. 38 (2021) 145020 (36pp)

\bibitem{ArgMa4}J. Argota-Quiroz and S. Majid, Quantum gravity on finite spacetimes and dynamical mass, in press Corfu Summer Institute 2021: School and Workshops on Elementary Particle Physics and Gravity, PoS (2022) (41pp)

 \bibitem{BegMa} E.J. Beggs and S. Majid, {\em Quantum Riemannian Geometry},  Grundlehren der mathematischen Wissenschaften, Vol. 355, Springer (2020) 809pp

 \bibitem{BegMa:spe}E.J. Beggs and S. Majid, Spectral triples from bimodule connections and Chern connections, J. Noncomm. Geom., 11 (2017) 669--701
 

\bibitem{ChaCon}A. Chamsedine and A. Connes, Why the Standard Model, J. Geom. Phys. 58 (2008) 38-47

\bibitem{Con}
A. Connes,   Noncommutative Geometry, 
Academic Press, Inc., San Diego, CA, 1994

\bibitem{ConMar} A. Connes and M. Marcolli, {\em Noncommutative Geometry, Quantum Fields and Motives} (AMS Colloquium Publications Vol 55), Hindustan Book Agency, 2008. 




\bibitem{DabSit} L. Dabrowski and A. Sitarz, Fermion masses, mass-mixing and the almost commutative geometry of the Standard Model, JHEP (2019) 68

\bibitem{Dev}A. Devastato, S. Farnsworth, F. Lizzi and P. Martinetti, Lorentz signature and twisted spectral triples, JHEP (2018) 89

\bibitem{DFR}S. Doplicher, K. Fredenhagen and J. E. Roberts, The quantum structure of spacetime at the Planck scale and quantum fields, Commun. Math. Phys. 172 (1995) 187--220


\bibitem{DVM}
M. Dubois-Violette and  P.W.\ Michor, Connections on central bimodules in 
noncommutative differential geometry, J.\ Geom.\ Phys.\ 20 (1996) 218--232


\bibitem{Hoo}G. 't Hooft, Quantization of point particles in 2+1 dimensional gravity and space- time discreteness, Class. Quant. Grav. 13 (1996) 1023

\bibitem{LirMa}E. Lira-Torres and S. Majid, Quantum gravity and Riemannian geometry on the fuzzy sphere, Lett. Math. Phys. (2021) 111:29 (21pp)

\bibitem{LirMa2}E. Lira-Torres and S. Majid, Geometric Dirac operator on the fuzzy sphere,  Lett. Math. Phys. (2022) 112:10

\bibitem{LirMa3} E. Lira-Torres and S. Majid, Geometric Dirac operator on noncommutative torus and $M_2(C)$,  Lett. Math. Phys. (2024) 114:70

\bibitem{LiuMa1}C. Liu and S. Majid, Quantum geodesics on quantum Minkowski spacetime,  J. Phys. A (2022)

\bibitem{LiuMa2}C. Liu and S. Majid, Quantum Kaluza-Klein theory with $M_2(C)$, J. High Energ. Phys. (2023) 102

\bibitem{LiuMa3}C. Liu and S. Majid, C. Liu and S. Majid, Yang-Mills fields from fuzzy sphere quantum Kaluza-Klein model, J. High Energ. Phys. 07 (2024) 195


\bibitem{Mad}J. Madore, The fuzzy sphere, Class. Quantum Grav. 9 (1992) 69--88


\bibitem{Ma:pla}S. Majid, Hopf algebras for physics at the Planck scale, Class. Quantum Grav. 5 (1988) 1587--1607

\bibitem{Ma:sq}S. Majid, Quantum gravity on a square graph, Class. Quantum Grav 36 (2019) 245009 (23pp) 

\bibitem{Ma:haw}S. Majid, Quantum Riemannian geometry and particle creation on the integer line, Class. Quantum Grav 36 (2019) 135011 (22pp)

\bibitem{Ma:dir}S. Majid, Dirac operator associated to a quantum metric, in press J. Noncomm. Geom. (2024)



\bibitem{Mou}
J.\ Mourad, Linear connections in noncommutative geometry, Class.\ Quant. 
Grav.\ 12 (1995)  965--974


\bibitem{KK}J.M. Overduin and P.S.  Wesson, Kaluza-Klein Gravity. Physics Reports. 283 (1997)  303--378

\bibitem{Sch} M.D. Schwarz, {\em Quantum field theory and the Standard Model}, Cambridge University Press (2014)

\bibitem{Sny}H.S. Snyder,  Quantized space-time Phys. Rev. 71 (1947) 38--41

\bibitem{KKhermitian}M. D. Pollock, On the Dirac equation in curved space-time, Acta Phys. Polon. B 41 (2010) 1827--1846

\bibitem{freedman}D. Z. Freedman and A. van Proeyen, Supergravity, Cambridge University Press, Cambridge, U.K. (2012)

\bibitem{Str}R.L. Stratonovich, Sov. Phys. JETP 31 (1956) 1012


 \end{thebibliography}
\end{document}